\documentclass[onecolumn,aps,prd,showpacs,preprintnumbers,groupedaddress]{revtex4}
\usepackage{amsmath}
\usepackage{graphicx}
\usepackage{bm}
\usepackage{color,colordvi}
\usepackage{dcolumn}
\begin{document}

\newcommand{\as}{\alpha_{\textrm s}}
\newcommand{\eeZA}{$e^+e^- \rightarrow Z^0 A^0$}
\newcommand{\eeZh}{$e^+e^- \rightarrow Z^0 h^0$}
\newcommand{\AgZ}{$A^0 \rightarrow \gamma Z^0$}
\newcommand{\AZZ}{$A^0 \rightarrow Z^0 Z^0$}

\preprint{YITP-SB-05-03}
\preprint{BNL-HET-05/4}
\preprint{hep-ph/0502195}
\title{Associated production of $A^0$ and $Z^0$ bosons and \\ 
       Rare Pseudoscalar Higgs Decays}
\author{B. Field}
\email[]{bfield@ic.sunysb.edu}
\affiliation{C.N. Yang Institute for Theoretical Physics, 
             Stony Brook University,
             Stony Brook, New York 11794-3840, USA}
\affiliation{Department of Physics, Brookhaven National Laboratory,
             Upton, New York 11973, USA}
\date{February 21, 2005}

\begin{abstract}

We study the production of a pseudoscalar Higgs boson $A^0$ in
association with a $Z^0$ boson at a future international linear collider
(ILC). We consider the contributions to this process at the one loop
level in the Minimal Supersymmetric Standard Model (MSSM) from top and
bottom quarks as well as stop and sbottom squarks. We also study the
squark contributions to the decay widths of the pseudoscalar Higgs boson
for the decays $A^0 \rightarrow \gamma Z^0$ and $A^0 \rightarrow Z^0
Z^0$. The contribution from the supersymmetric loops are found to be
directly proportional to the squark mixing and potentially large due to 
the massive pseudoscalar Higgs coupling to squarks.

\end{abstract}

\pacs{13.66.Fg, 14.70.Hp, 14.80.Cp}
\maketitle

\section{Introduction}

The Higgs mechanism is the means by which the electroweak symmetry is
broken in the Standard Model (SM) and in the Minimally Supersymmetric
Standard Model (MSSM) (see \cite{Gunion:1989we, Carena:2002es,
Heinemeyer:2004gx} for review). The MSSM has two Higgs doublets which
are used to generate the masses of the up- and down-type quarks. This
leads to five physical Higgs bosons. These consist of two CP-even
neutral scalar bosons ($h^0$, $H^0$), one CP-odd neutral pseudoscalar
($A^0$), and two charged bosons ($H^\pm$). The extended Higgs sector has
several new parameters that can be determined with the input of two
parameters, the mass of the pseudoscalar $M_{A^0}$ and $\tan\beta$ given
reasonable assumptions as to the size of the other supersymmetric
parameters in the theory. The parameter $\tan\beta \equiv v_u/v_d$ is
the ratio of the vacuum expectation values (\textsc{vev}s) of the up and
down sectors. We are interested in the phenomenology of the pseudoscalar
Higgs boson at a future international linear collider, in particular,
what the heavy squark contributions can tell us about the phenomenology.

There have been many studies of pseudoscalar Higgs boson
phenomenology\cite{Harlander:2002vv, Field:2002gt, Field:2002pb,
Ravindran:2002dc, Field:2003yy, Field:2004tt, Field:2004nc}. The
phenomenology of the pseudoscalar Higgs produced in association with a
$Z^0$ boson at a hadron collider (such as the Tevatron and the CERN LHC)
has also been studied\cite{Kao:1991xg, Yin:2002sq, Kao:2003jw,
Kao:2004vp, Li:2005qn}, as well as at a future international linear
collider (ILC)\cite{Barger:1993wt, Akeroyd:1999gu, Akeroyd:2001ak,
Farris:2002ny, Arhrib:2002ti}. There have been some studies of the
decays of Higgs bosons into squarks\cite{Bartl:1997yd, Eberl:1999he}
that show how large these processes can become, but squark contributions
are rarely taken into account for pseudoscalar Higgs processes. In this
paper we will discuss squark contributions to processes involving the
production and decay of a pseudoscalar Higgs boson which have been
missing from the literature thus far. 

The construction of an ILC would greatly enhance our ability to measure
the parameters in many new physics scenarios\cite{Accomando:1997wt,
Dawson:2004xz} including the MSSM. The process \eeZA\ at an ILC is
interesting because it has the possibility of four bottom quarks in the
final state allowing for precise measurements of the process as well as
many other excellent final states. This process is also interesting
because it could compete with the \eeZh\ process. The \eeZA\ process
occurs at the one loop level because the pseudoscalar Higgs boson does
not couple to vector bosons at tree level. This process also allows for
the exploration of squark mixing for the heaviest two generations of
squarks.

We find that the dominant contributions to the \eeZA\ process does not
come from top and bottom quark loops. We find that the squark
contributions to the \eeZA\ process are relevant at all values of
$\tan\beta$ and although they depend on the mixing in the stop and
sbottom squark sectors, they dominate over the standard model field
contributions due to the large coupling of the pseudoscalar to squarks. 
We find that the pseudoscalar decay \AZZ\ has significant contributions
from the squark sector and that the \AZZ\ branching ratio becomes more
important for large values of the mass of the pseudoscalar. We also find
that the \AgZ\ decay squark has contributions on the order of the quark
contributions for a light pseudoscalar Higgs at large values of
$\tan\beta$. Overall, squark contributions need to be added to processes
involving neutral Higgs bosons to complete our understanding of their
phenomenology.

\section{Squark Contributions}

In the SM, quarks generically couple to the pseudoscalar Higgs boson as
$\gamma^5 m_q/v$. Beyond this base coupling, the up- and down-type
quarks couple to the pseudoscalar differently. Up-type quarks couple as
$\cot\beta$ and down-type quarks couple as $\tan\beta$, leading to the
well-known conclusion that bottom quarks become more important at large
values of $\tan\beta$. This is also true in the stop/sbottom sectors. 
The sbottom squarks become more important as $\tan\beta$ becomes large.

In the MSSM, right and left handed quarks, $q_{R,L}$, have scalar
super-partners, $\tilde{q}_{L,R}$. We are interested in the stop and
sbottom squarks ($\tilde{t}_R, \tilde{t}_L, \tilde{b}_R, \tilde{b}_L$).
The squark sector of the MSSM has the possibility for squarks to mix
into mass eigenstates that are different than the usual left-right
basis. We can introduce a mixing angle that diagonalizes the squarks
into their mass eigenstates. This can be written generically for the new
squark eigenstates $\tilde{q}_{1,2}$ as\cite{Haber:1984rc}
\begin{align}
\label{eqn:mix} \nonumber
\tilde{q}_1 &= \phantom{-}\tilde{q}_L c_q + \tilde{q}_R s_q, \\ 
\tilde{q}_2 &=          - \tilde{q}_L s_q + \tilde{q}_R c_q,
\end{align}
where $c_q \equiv \cos\theta_q$ and $s_q \equiv \sin\theta_q$ are the
mixing angles in each of the squark sectors. Squark mixing is
particularly interesting when it comes to the squark loop contributions
for pseudoscalar Higgs production. In the left-right basis, the
pseudoscalar Higgs couples only to squarks that change their handedness
at the vertex leaving only vertices such as $A^0 \tilde{q}_R
\tilde{q}_L$. However it is now easy to see that in the presence of
squark mixing, vertices of the type $A^0 \tilde{q}_i \tilde{q}_j$ for
all values of $(i,j)$ are present and retain the size of the $A^0
\tilde{q}_R \tilde{q}_L$ vertex times a function of the new mixing
angle. This is also true of other neutral Higgs bosons in the MSSM, but
the analysis is more complicated as there are more vertices present for
the $h^0$ and $H^0$ Higgs bosons.

In this model, all of the pseudoscalar Higgs vertices in the squark
sector are proportional to the squark mixing angles. Thus the mixing
angles regulate the squark contribution to any pseudoscalar Higgs
process. It is possible to tune the mixing angles $s_q$ and $c_q$ to
eliminate all but one of these vertices. For instance if
$\sin\theta_t=0$, then only the $A^0 \tilde{t}_1 \tilde{t}_2$ vertex
remains and it is maximal. It is easy to see in Eqn.~(\ref{eqn:mix})
that this would correspond to being back in the left-right basis.

The mixing angles themselves are not completely independent from the 
other parameters in the MSSM. We can write the mixing angles as
\begin{align}
\sin (2\theta_t) &= \frac{2 m_{\textrm{top}} (A_t - \mu \cot\beta)}
                         {m^2_{\tilde{t}_1}-m^2_{\tilde{t}_2}}, \\
\sin (2\theta_b) &= \frac{2 m_{\textrm{bot}} (A_b - \mu \tan\beta)}
                         {m^2_{\tilde{b}_1}-m^2_{\tilde{b}_2}}.
\end{align}
Here $A_{t,b}$ are the tri-linear scalar couplings for soft
supersymmetry breaking in the MSSM Lagrangian and $\mu$ is the
supersymmetry breaking Higgs mass parameter in the super potential.

We can see that there is more splitting in the stop sector as it is
proportional to the top quark mass which is much larger than the bottom
quark mass. This is a general feature of squark mixing. In our analysis
we picked $\tilde{m}_1 \equiv 1$~TeV in both the stop and sbottom
sectors and generated $\tilde{m}_2$ using the other parameters in the
theory for an arbitrary mixing angle allowing us to study the mixing in
the squark sector. Thus we always have $\tilde{m}_2 > \tilde{m}_1$ in
the stop sector. This is the standard ordering for squark masses. Our
choices for squark masses are well beyond current experimental
limits\cite{Eidelman:2004wy} and we have chosen such heavy squarks to
make our results more conservative. The other parameters in the theory
were chosen to be $\mu = 300$~GeV and $A_{t,b}=1500$~GeV. Of course, the
squark contributions to the process depend on the mass of the squarks
and lighter squarks lead to larger contributions to any process
involving them. In choosing $\tilde{m}_1 = 1$~TeV we find maximal mixing
in the squark sector leads to the lightest masses for the second quark.
Although, our results show a very complicated dependence on these mixing
angles, the dominant effect of the supersymmetric contributions are seen
when the squarks are the lightest as is to be expected.

These mass relations are tree level relations and the corrections to the
mixing angles are known in the literature\cite{Carena:1999py}. However,
since we are interested in studying the effects of the squark mixing on
our observables, it is unclear how to incorporate these corrections to
produce squark masses that obey specific mixing angles. The effects of
these higher order corrections are the most pronounced in the sbottom
sector and future studies should be completed on this subject. We have
used the $\overline{\textsc{ms}}$ running-mass of the bottom quark in
our generation of the second sbottom squark mass.

\begin{table}
\begin{tabular}{|c|c|cccc|}
\multicolumn{6}{c}{$\Phi\tilde{q}_i\tilde{q}_j$ Coupling} \\
\hline \hline
              &          & \multicolumn{4}{c|}{$(i,j)$} \\
 $\Phi$       & coupling & 11 & 12 & 21 & 22 \\
\hline
$\gamma^\mu$  & $- i e Q_q (p\!+\!q)^\mu$ 
              & $1$
              & $0$
              & $0$
              & $1$ \\
$Z^\mu$       & $-(ie/s_w c_w)(p\!+\!q)^\mu$
              & $[T^3_q c_q^2 - Q_q s_w^2]$
              & $-T^3_q s_q c_q$
              & $-T^3_q s_q c_q$
              & $[T^3_q s_q^2 - Q_q s_w^2]$ \\
$Z^\mu Z^\nu$ & $(2ie^2 / s_w^2 c_w^2) \eta^{\mu\nu}$
              & $[T^3_q c_q^2 - Q_q s_w^2]$
              & $-T^3_q s_q c_q$
              & $-T^3_q s_q c_q$
              & $[T^3_q s_q^2 - Q_q s_w^2]$ \\
$A^0$         & $\tilde{A}_q$
              & $c_q s_q$ 
              & $c_q^2$ 
              & $- s_q^2$ 
              & $- c_q s_q$ \\
\hline
\end{tabular}

\caption{Feynman rules for squark mass eigenstate couplings to the
photon, a $Z^0$ boson, two $Z^0$ bosons, and the pseudoscalar Higgs
boson. Here the squarks have momentum $p$ and $q$ (running in the
direction of the flow of charge) and the photon and $Z^0$ boson have a
free Lorentz index $\mu$. The two $Z^0$ bosons have free Lorentz indices
$(\mu,\nu)$. The $(i,j)$ indices are written in the direction of the
flow of charge at the vertex.  $T^3_q$ is the isospin of the left-handed
quark for which the squark is the super-partner, $\pm \frac{1}{2}$,
$Q_q$ is the charge of the squark, $s_w \equiv \sin\theta_w$, and $c_w
\equiv \cos\theta_w$. The photon couples diagonally to the mixed
squarks. The pseudoscalar coupling constant is listed in the text and is
potentially very large compared to the coupling in the quark sector.}

\label{tbl:feynman}
\end{table}

\section{Pseudoscalar Coupling to Squarks}

Squark mixing imposes new Feynman rules for the vertices in our
processes. The results of the application of the squark mixing to the
pseudoscalar vertices can be found in Table~\ref{tbl:feynman}. The
coupling of the squarks to the pseudoscalar can be written as
\begin{align}
\tilde{A}_t &= -\frac{m_{\textrm{top}}}{v} ( \mu - A_t\cot\beta ), \\ 
\tilde{A}_b &= -\frac{m_{\textrm{bot}}}{v} ( \mu - A_b\tan\beta ), 
\end{align}
where $v=\sqrt{v_1^2+v_2^2}=246$~GeV is the Higgs \textsc{vev} and we
have chosen sign$(\mu) <0$ implicitly in our choice of couplings.

The squark contributions to pseudoscalar Higgs processes are often
missing in the literature and were believed to be quite small because
they are loop suppressed and the squark would be very heavy if they
exist in nature. Although this is certainly a possibility given the
right parameter choice, the pseudoscalar coupling to squarks is quite
large and is not entirely suppressed by the heavy squark loops. This is
due to the fact that the pseudoscalar couples to right- and left-handed
squarks only in a non-diagonal way. In the MSSM, the non-diagonal squark
vertices with any neutral Higgs boson ($\Phi = \{ h^0, H^0, A^0 \}$),
written here as $\Phi \tilde{q}_R \tilde{\bar{q}}_L$, is proportional to
the tri-linear soft-supersymmetry breaking parameter multiplied by the
same factor as the coupling in the quark sector modulo any mixing angles
that may also be present at the vertex. For the pseudoscalar Higgs,
there are no mixing angles present, just $\tan\beta$ or $\cot\beta$
depending on the nature of the squark. This could also increase this
effect in the sbottom sector once the difference in the masses between
the top and bottom quarks is overcome. Therefore the squark-squark-Higgs
coupling is potentially greatly enhanced compared to the
quark-quark-Higgs coupling, so much so as to overcome the loop
suppression and the heavy squark masses to become the dominant
contribution.

If we consider the top/stop sector as an illustration, the importance of
the squark contributions will become clear. The vertex $A^0 t \bar{t}
\sim m_t/v$ and the dominant contribution to the vertex $A^0 \tilde{t}
\tilde{\bar{t}} \sim m_t/v \times A_t \cot\beta$. Thus, the squark
vertex is as large at the quark vertex times a large factor, in our case
$1500$~GeV for $\tan\beta=1$. This factor of $10^3$ is squared in a
cross-section or width and gives an overall factor of $10^6$. We have
picked squarks with masses on the order of a TeV (about five times the
mass of the top quark) which deceases the cross-section or width by an
order of magnitude when compared to a squark with a mass lowered to that
of the top quark. So na\"{\i}vely we would consider the squark
contributions to be larger than the quark contributions by a factor of
$10^5$. This is exactly what we have found for the \eeZA\ process and
the \AZZ\ width. There are some cancellations in the \AgZ\ width leading
to a result that is on the same order as the contributions from the SM
fields for some values of $M_{A^0}$ and $\tan\beta$.

We would like to emphasize that this is the case for the non-diagonal
squark vertices in the entire neutral Higgs sector of the MSSM. The $h^0
\tilde{q}_R \tilde{\bar{q}}_L$ and $H^0 \tilde{q}_R \tilde{\bar{q}}_L$
should show similar enhancements with some additional dependence on the
mixing angles $(\alpha,\beta)$. The squark loop processes involving the
production of these particles could very well be larger than the
tree-level production processes and could lead to better bounds on
squark masses or trilinear soft-supersymmetry breaking terms from the
LEP2 data for the production of the other neutral Higgs bosons in the 
MSSM.

It should be noted that this is partially an arbitrary enhancement
because we have chosen the soft-supersymmetry breaking parameters to be
large. However, if the soft breaking terms exist in nature, they will
enhance the squark-squark-Higgs vertex over that of the
quark-quark-Higgs vertex proportionately to their size. With this
information we are ready to construct the matrix elements for our decays
\AgZ\ and \AZZ\ and our process \eeZA.

\section{Matrix Elements and Results}

\begin{figure}
  \begin{tabular}{ccc}
  \resizebox{50mm}{!}{\includegraphics{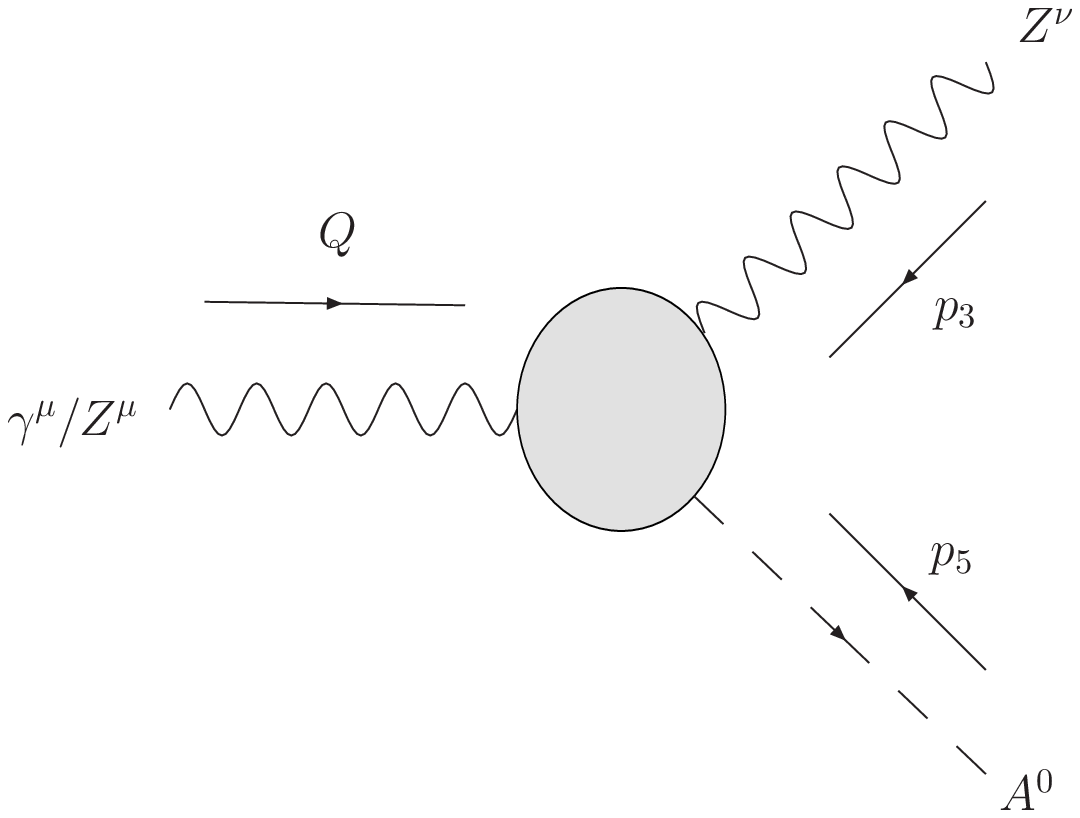}}   &
  \resizebox{55mm}{!}{\includegraphics{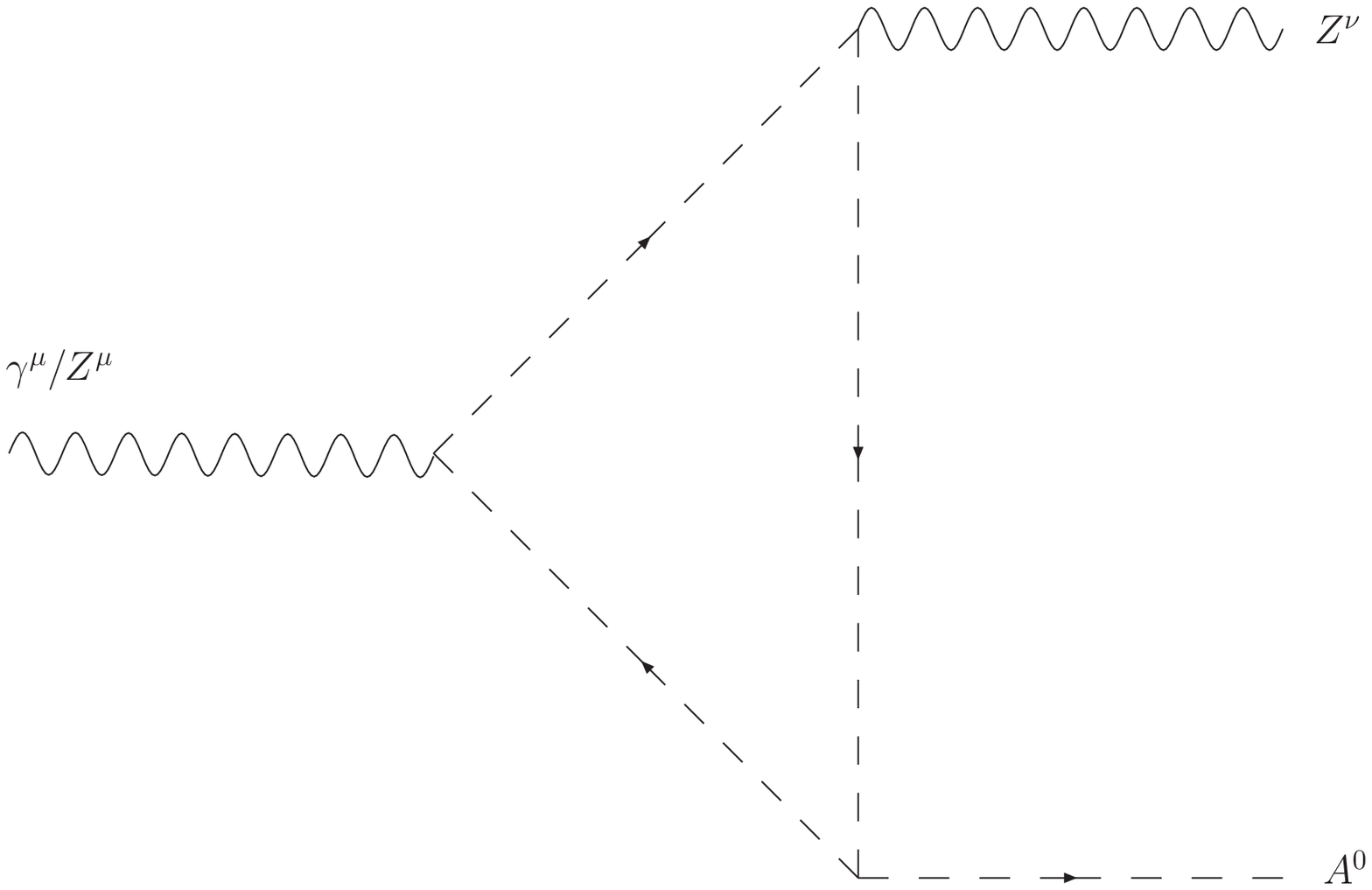}} &
  \resizebox{55mm}{!}{\includegraphics{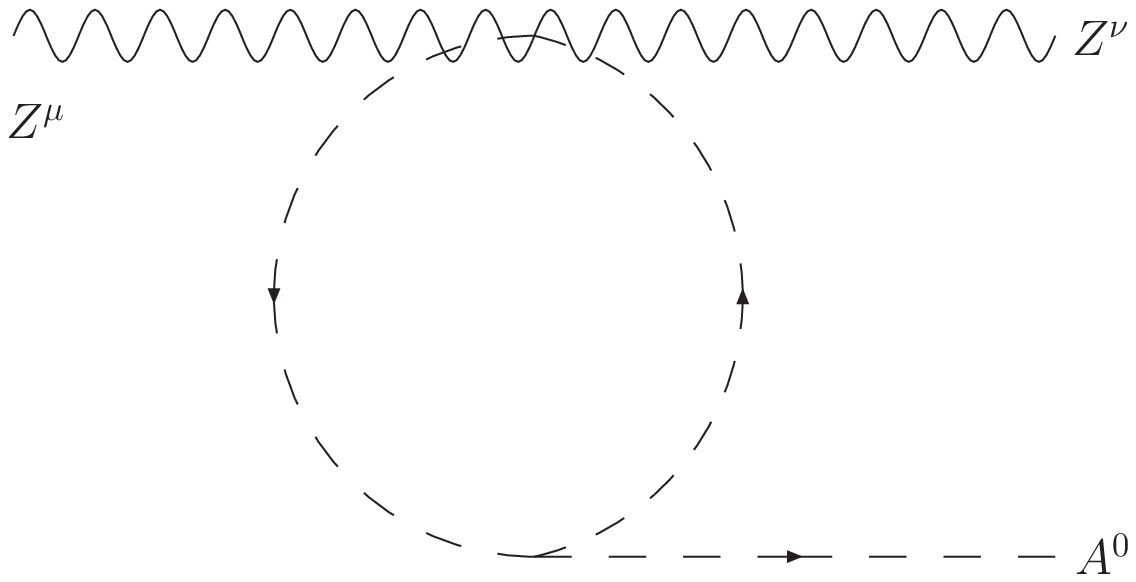}}
  \end{tabular}
\caption{The three-point functions needed for the \eeZA\ calculation. 
Top and bottom quarks as well as stop and sbottom squarks were allowed 
to run inside the blob. The momentum assignment was chosen to make the 
\eeZA\ calculation simpler.}
\label{fig:decay}
\end{figure}

To understand the \eeZA\ process, we calculated the three point
functions $\gamma Z^0 A^0$ and $Z^0 Z^0 A^0$ with both top and bottom
quarks and stop and sbottom squarks. A generic representation of the
three-point functions can be seen in Fig.~(\ref{fig:decay}). The
contributions from the top and bottom quarks were checked against known
results\cite{Gunion:1991cw}. We also checked our calculation with the
known partial width for $\Gamma_i(A^0 \rightarrow \gamma Z^0)$ as given
by \textsc{hdecay}\cite{Djouadi:1997yw}. We found excellent agreement.
We also found that due to the tensor structure of the quark and squark
contributions, the two processes do not interfere with each other. The
exact form of the contributions is worked out in the appendix and it is
shown explicitly that the two contributions do not mix.

The squark contributions to $\Gamma_i(A^0 \rightarrow \gamma Z^0)$
become less and less important as $M_{A^0}$ becomes large. This can be
easily seen in our parameterization of the matrix elements squared 
below
\begin{equation}
\label{eqn:agz}
|\mathcal{M}^{\textsc{susy}}|^2_{A^0 \rightarrow \gamma Z^0} = 
\frac{\alpha^2 N_c^2}{16 \pi^2} \frac{1}{(16\pi^2)^2} \biggl\{
3 \biggl| \sum_q A^q_\gamma \biggr|^2 - 
   \frac{(M_{A^0}^2-M_Z^2)^2}{4} 
  \biggl| \sum_q E^q_\gamma \biggr|^2
\biggr\},
\end{equation}
where the superscript \textsc{susy} here implies that this is the 
contribution from the squark fields (not the SM fields) and $N_c=3$ 
is the number of colors.

The functional form of the $A^q_\gamma, E^q_\gamma$ functions can be
found in the appendix. However, it is simple to see in
Eqn.~(\ref{eqn:agz}) that the second term makes the amplitude smaller
and smaller as $M_{A^0} \gg M_Z$. The functional form of the
$A^q_\gamma,E^q_\gamma$ functions guarantee the matrix elements are
positive definite. What cannot be immediately seen is that the two
pieces of the squark contribution almost cancel. Even though the
pseudoscalar squark vertices are greatly enhanced compared to their
standard model counterparts due to the trilinear term, the
supersymmetric contribution to the width is only slightly larger than
the SM contribution in some of the parameter space.

We also needed to calculate the three point function for $Z^0 Z^0 A^0$
for our \eeZA\ process. Although \textsc{hdecay} does not have this
final state for the pseudoscalar Higgs for comparison, we did calculate
the branching ratio $\Gamma_i(A^0 \rightarrow Z^0 Z^0)$ and added it to
our analysis. The matrix elements squared, listed below, have the
opposite effect from the $\gamma Z^0$ channel as they become more
important as the mass of the pseudoscalar grows. The matrix elements
grow like positive powers of $(M_{A^0} - 2M_Z)^n$. This can be seen in
the positive powers of this mass difference in the first and third terms
of the matrix elements,
\begin{align} \nonumber
|\mathcal{M}^{\textsc{susy}}|^2_{A^0 \rightarrow Z^0 Z^0} = 
\frac{\alpha^2 N_c^2}{32 \pi^2} \frac{1}{(16\pi^2)^2} \biggl\{ &
   \biggl( 2 + \frac{(M_{A^0}^2-2M_Z^2)^2}{4M_Z^4} \biggr) 
   \biggl| \sum_q A^q_Z \biggr|^2 \\ \nonumber
+& \biggl( M_Z^4 - \frac{1}{2}(M_{A^0}^2-2M_Z^2)^2 \biggr)
   \biggl| \sum_q E^q_Z \biggr|^2 \\
+& \biggr( \frac{(M_{A^0}^2-2M_Z^2)^3}{8M_Z^4} 
         -\frac{M_{A^0}^2-2M_Z^2}{2}
  \biggl)
  2 \, \textrm{Real} \biggl( \sum_q A^q_Z (E^q_Z)^\star \biggr)
\biggr\}.
\end{align}

\begin{figure}
\begin{tabular}{cc}
  \resizebox{75mm}{!}{\includegraphics{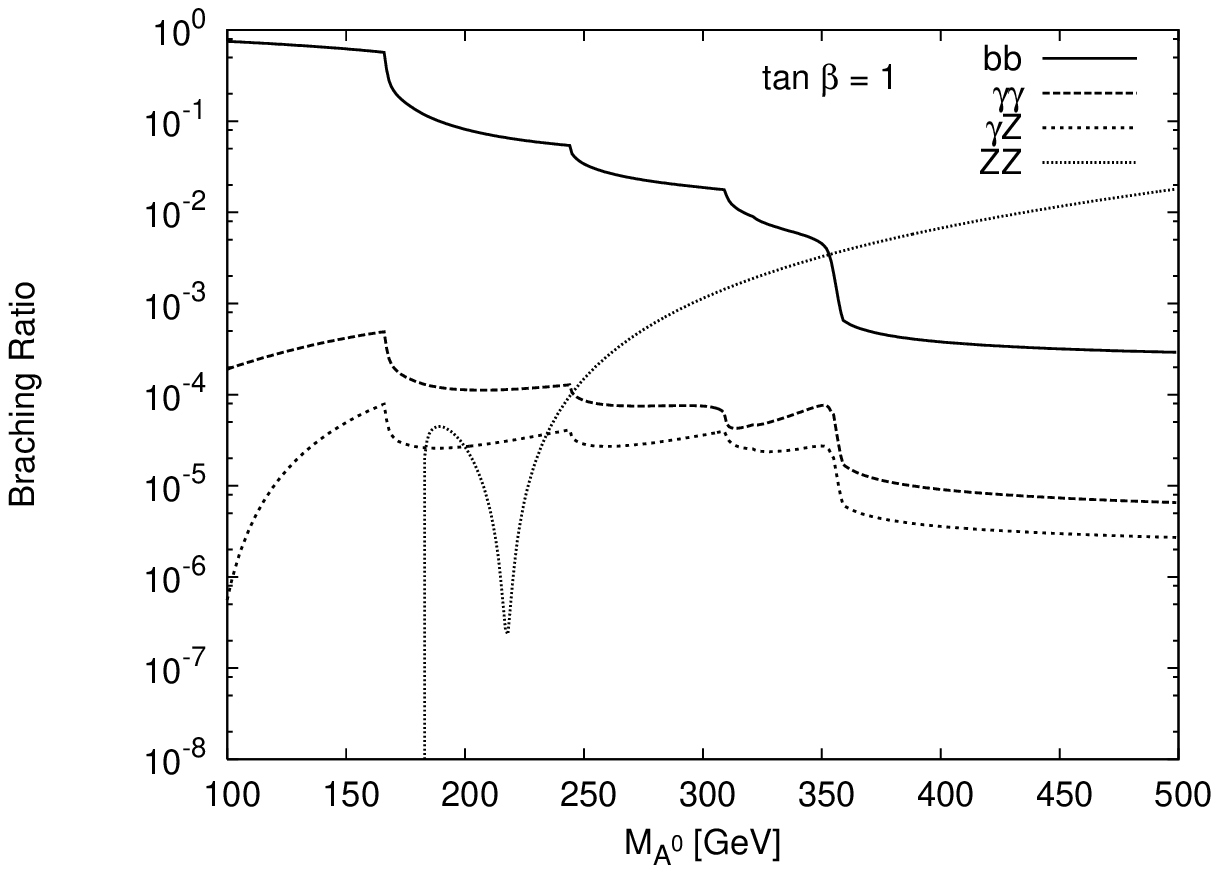}}  &
  \resizebox{75mm}{!}{\includegraphics{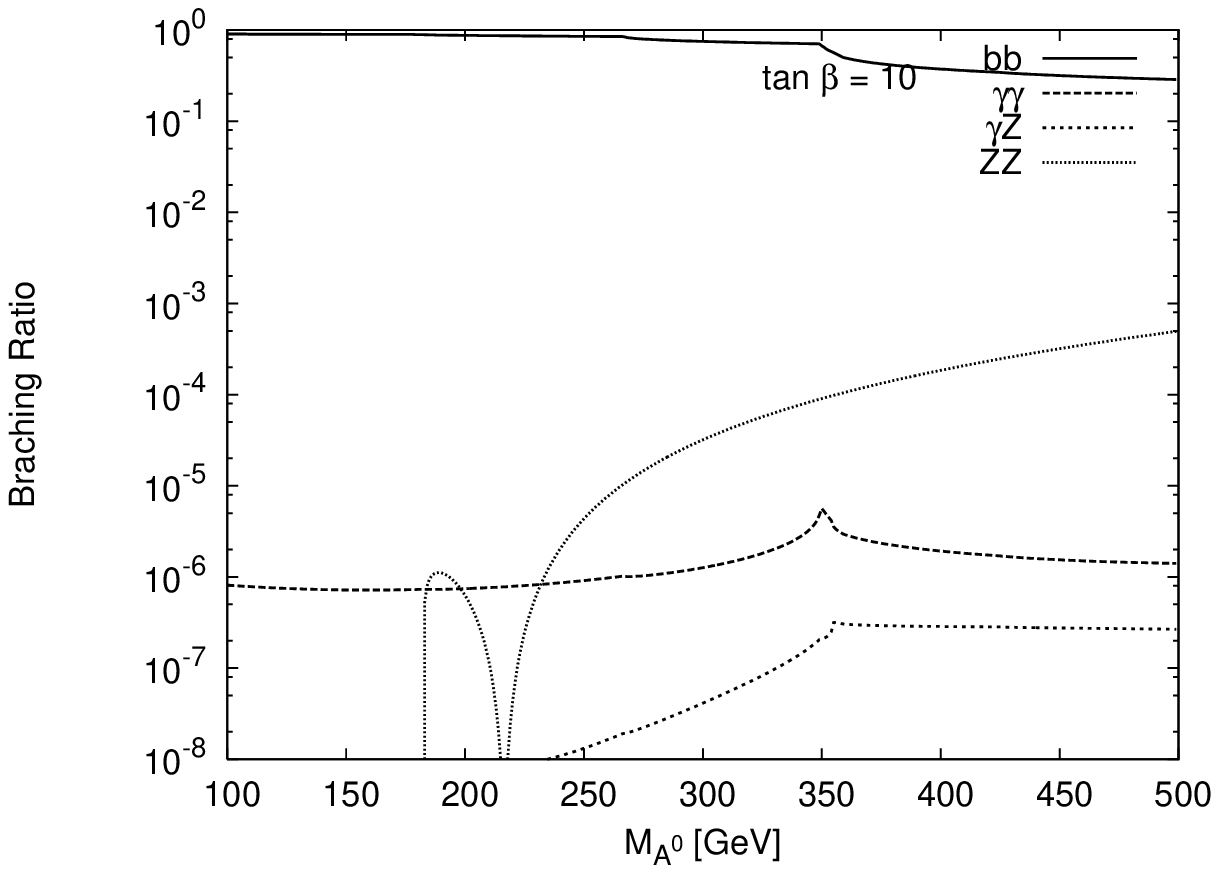}} \\
  \resizebox{75mm}{!}{\includegraphics{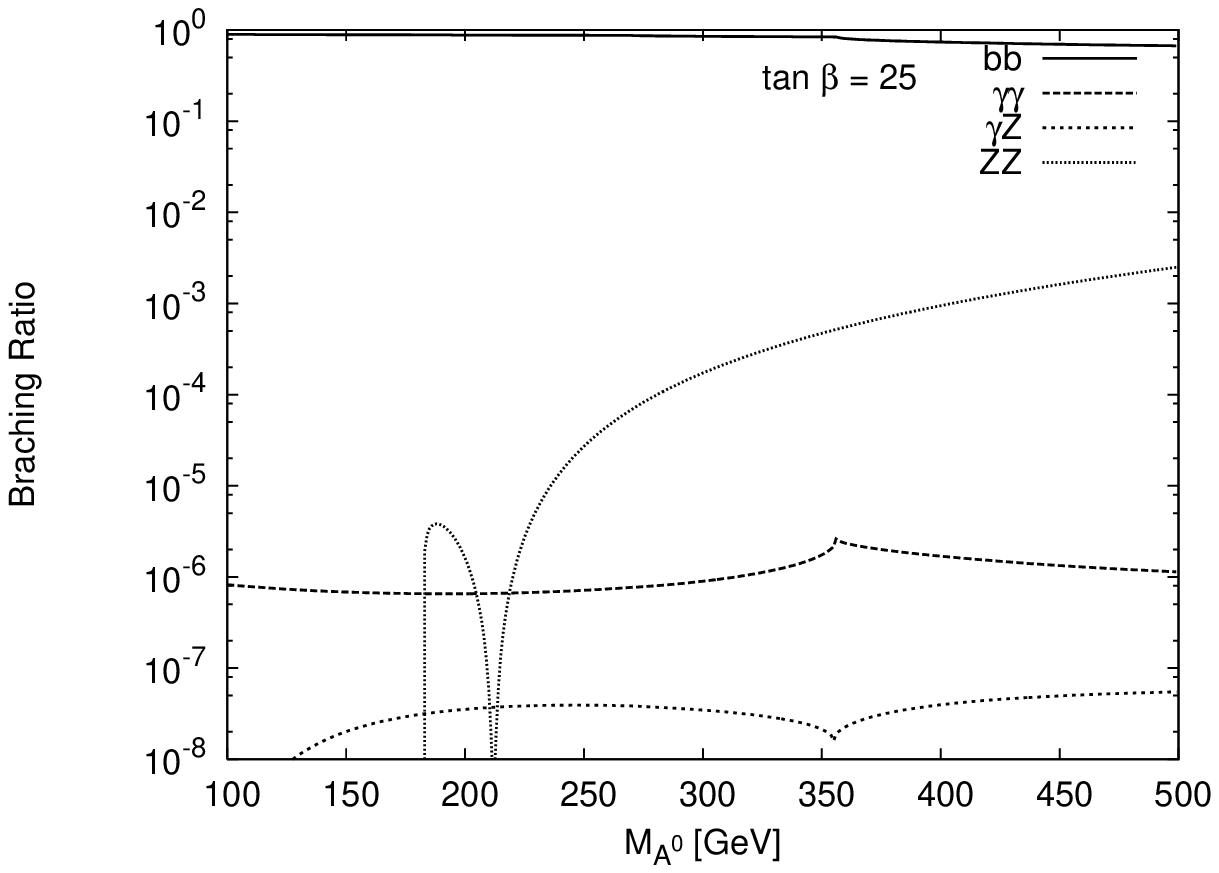}} &
  \resizebox{75mm}{!}{\includegraphics{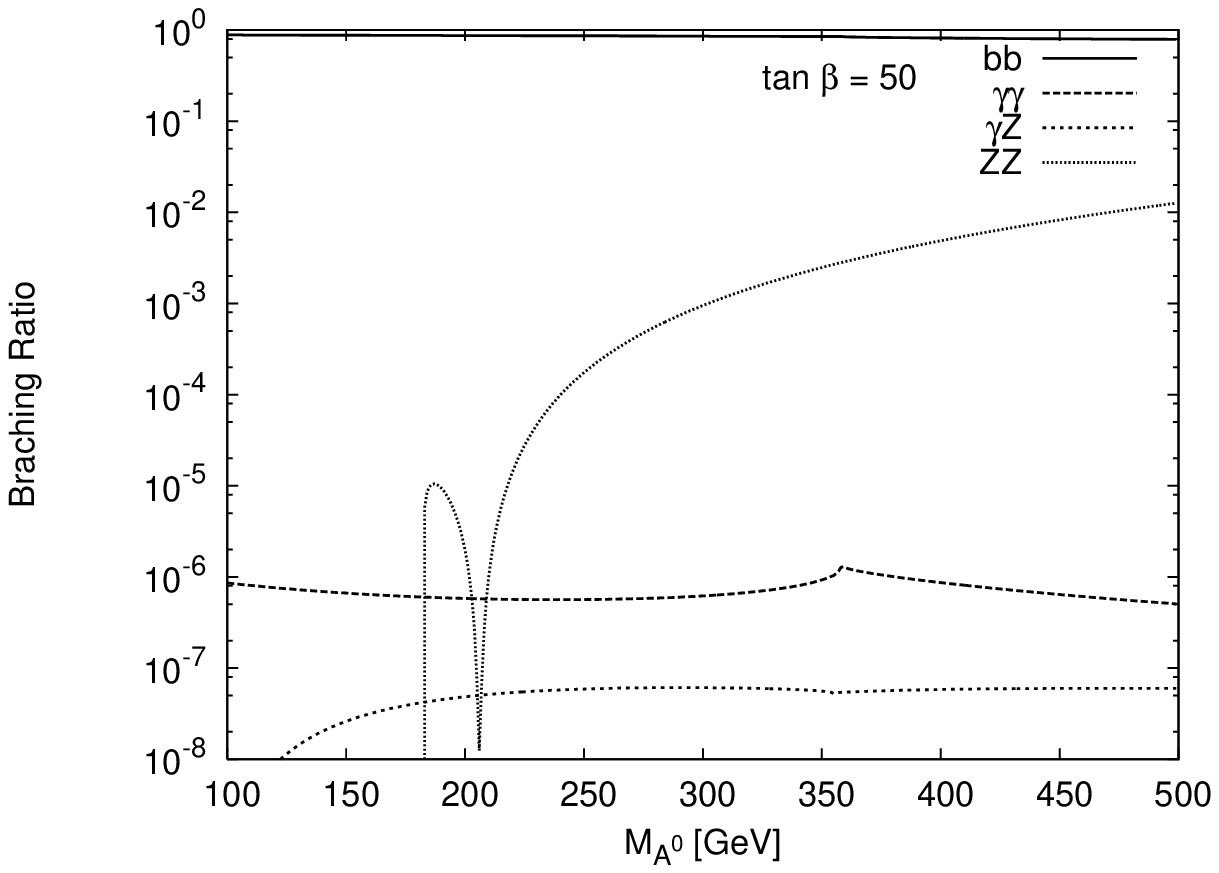}} \\
\end{tabular}

\caption{Improved branching ratios for the pseudoscalar Higgs boson
(including the $Z^0Z^0$ channel). The top and bottom quark loops as well
as stop and sbottom squark loops have been included in the \AgZ\ and
\AZZ\ channels. The other branching ratios have been taken from
\textsc{hdecay} and have been adjusted to allow for the new channel. We
see that the $b\bar{b}$ channel dominates at large $\tan\beta$, but the
$Z^0 Z^0$ channel also plays an important role. The dip in the $Z^0 Z^0$
channel is not a kinematic one, nor is it due to the opening of a new
decay channel in this case (the $W^+W^-$ channel has not been included). 
It is due to a value of the pseudoscalar Higgs mass that allows for
large cancellations to occur.}

\label{fig:gamma}
\end{figure}

The results of these partial widths can be seen in
Fig.~(\ref{fig:gamma}). These graphs were created using the output of
\textsc{hdecay} and adjusted to include the new contributions from the
squark loops in the \AgZ\ channel as well as the entirely new \AZZ\ 
channel.

\begin{figure}
  \begin{tabular}{cc}
  \resizebox{55mm}{!}{\includegraphics{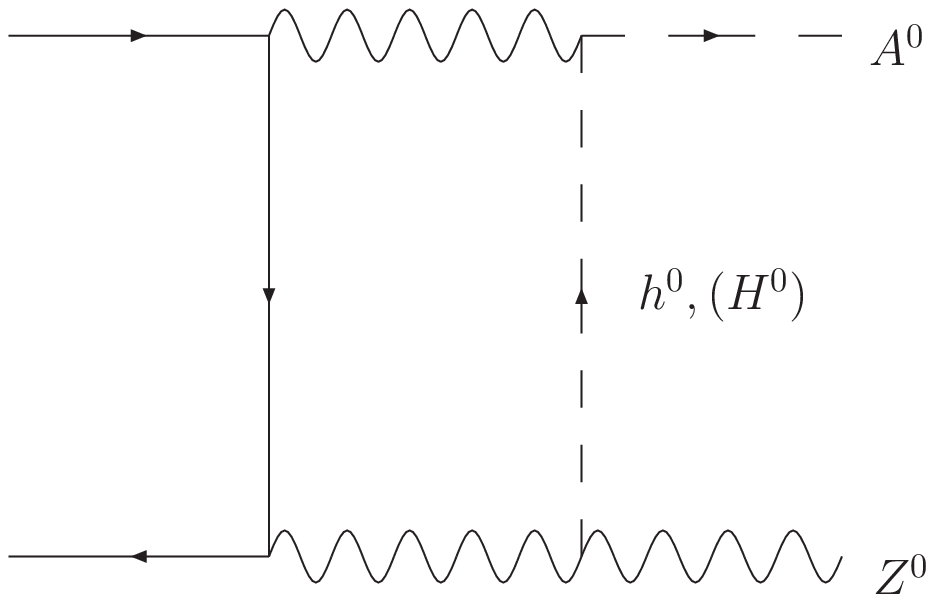}}  &
  \resizebox{60mm}{!}{\includegraphics{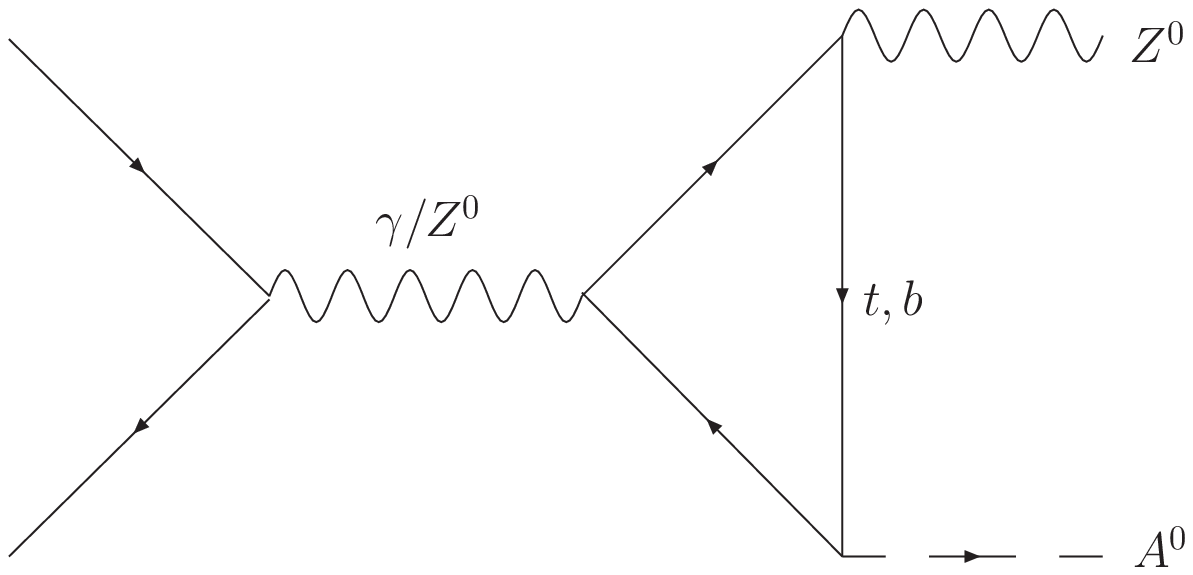}}   \\
  \resizebox{55mm}{!}{\includegraphics{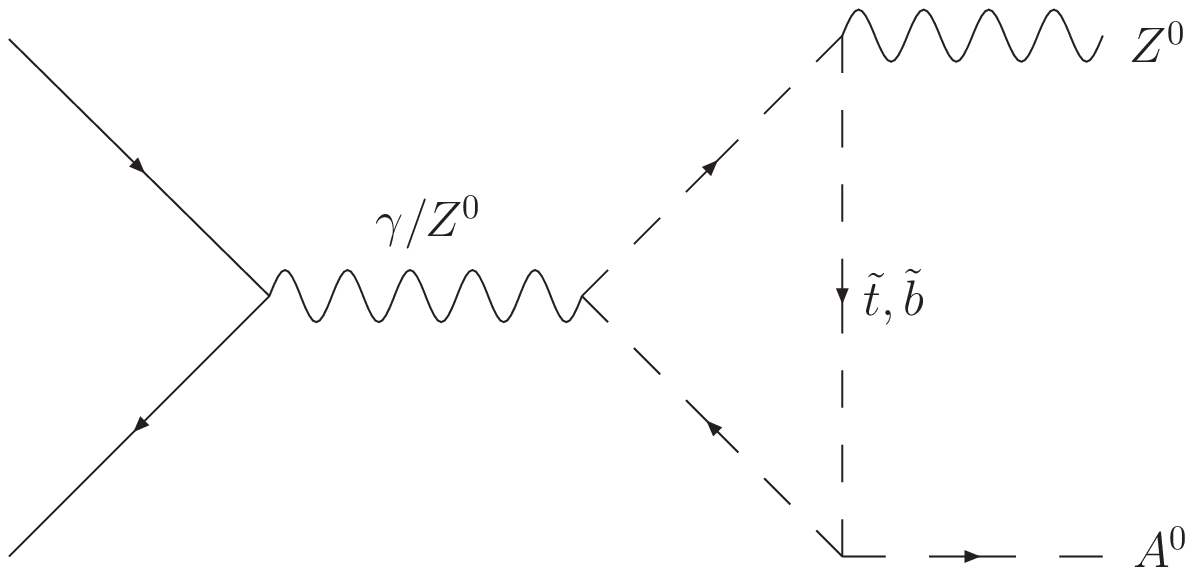}} &
  \resizebox{60mm}{!}{\includegraphics{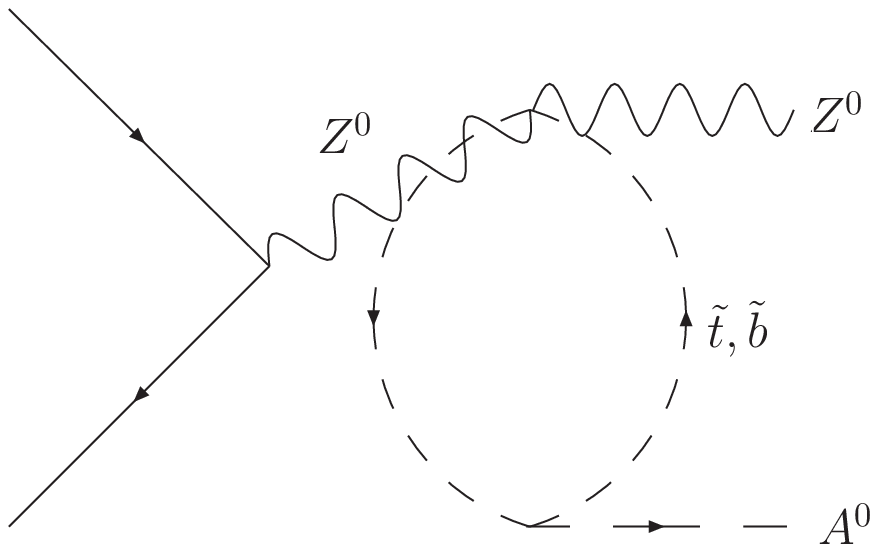}} 
  \end{tabular}

\caption{Diagrams contributing to the \eeZA\ process. The graphs with
the top and bottom quarks in the loop and the box graph are referred to
as the standard model contributions in the text. The squark loop graphs
are the dominant contribution to the processes.}

\label{fig:eeZA}
\end{figure}

Using our expressions for the two three-point functions with squark
loops, the squark contributions to the matrix elements for the \eeZA\
process can be written
\begin{align} \nonumber
\overline{|\mathcal{M}|}^2(e^+e^- \rightarrow Z^0 A^0) = 
32 \alpha^3 \pi^3 N_c^2 \frac{(tu-M_Z^2M_{A^0}^2)}{M_Z^2}
\biggl\{ &
\biggl( 1 + \frac{2sM_Z^2}{(tu-M_Z^2M_{A^0}^2)} \biggr) 
  \biggl( |x_1|^2 + |x_3|^2 \biggr) \\ \nonumber
+&\biggl( \frac{(t+u)^2}{4} - M_Z^2 M_{A^0}^2 \biggr)
  \biggl( |x_2|^2 + |x_4|^2 \biggr) \\ 
+&\biggl( \frac{(t+u)}{2} - M_Z^2 \biggr)
  2 \, \textrm{Real}\biggl( x_1x_2^\star + x_3x_4^\star \biggr)
\biggr\},
\end{align}
where,
\begin{align}
x_1 &= \sum_q \frac{A^q_\gamma}{s} + \frac{1}{4c_ws_w}
                                     \frac{1-4s_w^2}{s-M_Z^2} A^q_Z, \\
x_2 &= \sum_q \frac{E^q_\gamma}{s} + \frac{1}{4c_ws_w}
                                     \frac{1-4s_w^2}{s-M_Z^2} E^q_Z,  \\
x_3 &= - \sum_q \frac{1}{4c_ws_w}\frac{1}{s-M_Z^2}A^q_Z, \\
x_4 &= - \sum_q \frac{1}{4c_ws_w}\frac{1}{s-M_Z^2}E^q_Z.
\end{align}

The diagrams contributing to the \eeZA\ process are shown in
Fig.~(\ref{fig:eeZA}). The process was written as $e^+(p_1) e^-(p_2)
\rightarrow Z^0(-p_3) A^0(-p_5)$ and we have employed the usual
kinematic variables $s=(p_1+p_2)^2$, $t=(p_2+p_3)^2$, and
$u=(p_3+p_1)^2$. We took the electrons to be massless and set $p_1^2 =
p_2^2 = 0$, $p_3^2=M_Z^2$, and $p_5^2=M_{A^0}^2$.

\begin{figure}
\begin{tabular}{cc}
  \resizebox{75mm}{!}{\includegraphics{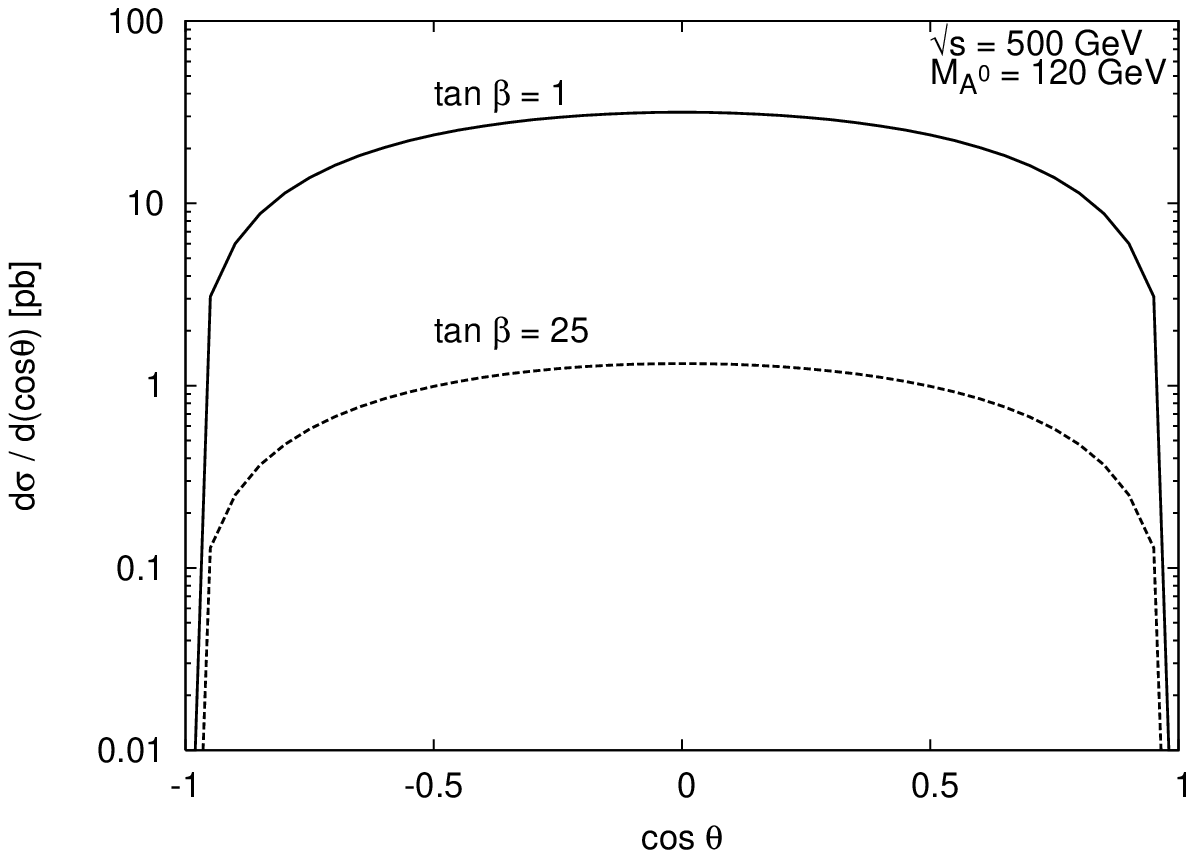}}  & 
  \resizebox{75mm}{!}{\includegraphics{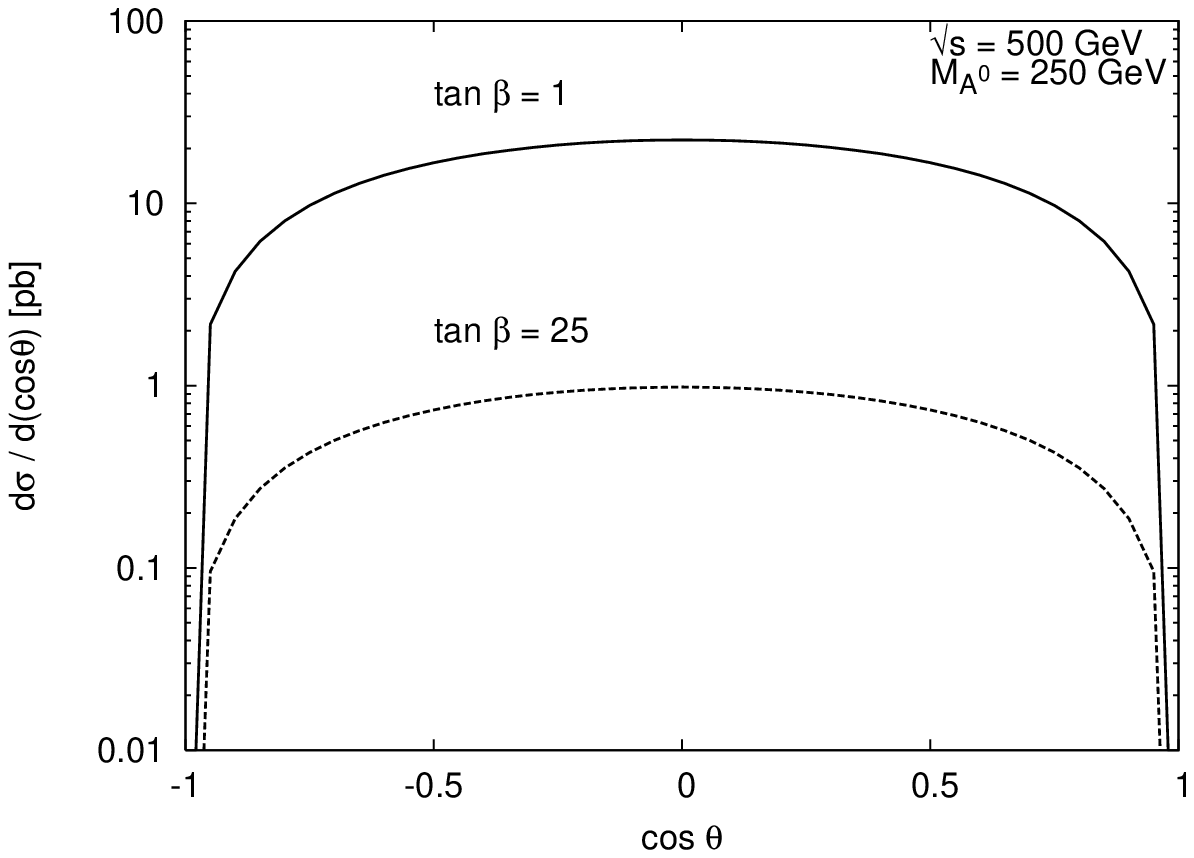}} \\
  \resizebox{75mm}{!}{\includegraphics{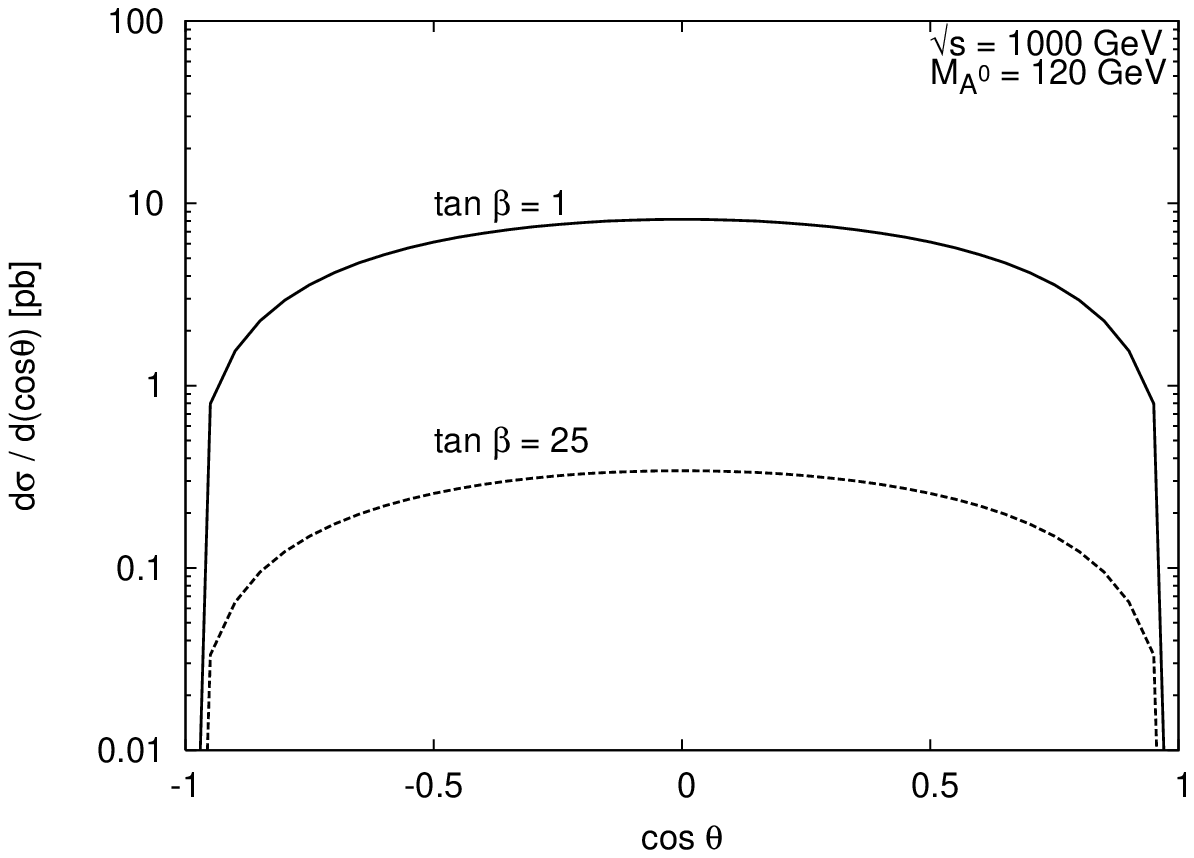}} & 
  \resizebox{75mm}{!}{\includegraphics{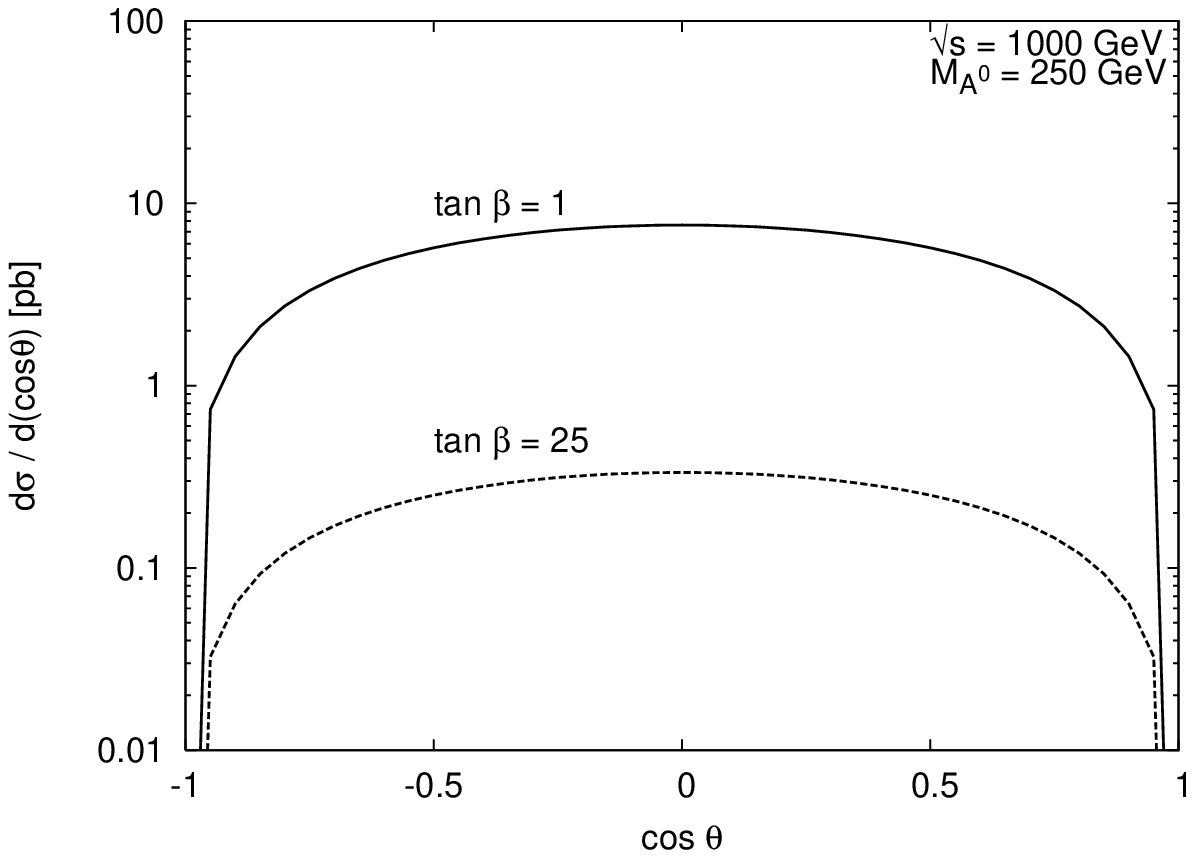}} 
\end{tabular}

\caption{Differential cross-sections for different parameters and
center-of-mass energies at a future international linear collider for
the process \eeZA. All are for the case of maximal mixing in each of the
squark sectors ($\sin(2\theta_{t,b})=1$) which leads to the largest
contributions. Only the contributions from the squark loops are shown as
they are several orders of magnitude greater than the contributions from
the SM fields which are on the order of $0.2$~fb for $M_{A^0}=120$~GeV
and $\tan\beta=1$. Increasing the mass of the pseudoscalar drops the
differential cross-section by about $30$\% and increasing the
center-of-mass energy to $1$~TeV decreases the differential
cross-section by about $75$\% for $\tan\beta=1$.}

\label{fig:sigma} 
\end{figure}

The differential cross-sections for a future ILC are shown in
Fig.~(\ref{fig:sigma}). The squark contributions can be seen for the two
different center-of-mass energies ($500$~GeV and $1$~TeV). The quark
contributions are not shown because they are several orders of magnitude
smaller than the squark contributions which are on the order of $0.2$~fb
for $M_{A^0}=120$~GeV and $\tan\beta=1$. The contributions from the SM
fields and the MSSM fields would not have mixed in any of the processes
in this paper due to their tensor structure, as described in the
appendix. The differential cross-section is symmetric with respect to
the scattering angle $\cos\theta$ and with the interchange of the
kinematic variables $t \leftrightarrow u$. 

\section{Conclusions}

In conclusion, we have found that the dominant contributions to the
\eeZA\ process does not come from top and bottom quark loops. The
extreme enhancement of the pseudoscalar Higgs-squark-squark vertices due
to the trilinear soft-supersymmetry breaking terms make the squark loops
the dominant contribution to this process. This result is generically
true in the entire neutral Higgs sector of the MSSM. Based on these
results, squark loops appear to play a very large roll in Higgs
phenomenology in the neutral sector.

\begin{acknowledgments}

The author is supported in part by the National Science Foundation grant
PHY-0098527 and under DOE Contract No. DE-AC02-98CH10886. The author
would like to thank S.~Dawson and J.~Smith for all their help and
comments as well as A.~Field-Pollatou and N.~Christensen for many other
helpful comments and suggestions.

\end{acknowledgments}

\appendix
\section{Matrix Elements}

The matrix elements for the process \eeZA\ has implicit in it two
one-loop three-point functions $\gamma Z^0 A^0$ and $Z^0 Z^0 A^0$. We
constructed these two three-point functions with quark and squark
contributions for the collider process and to understand the squark
contributions to our decay widths. There are also four box-type quark
diagrams that have not been discussed so far and are not related to the
three-point functions analyzed in this paper. These were included in our
analysis but they are not part of the squark sectors for which explicit
results will be presented and are negligible in size to the SM 
contributions.

The matrix elements for the rare pseudoscalar decays $A^0 \rightarrow
\gamma Z^0$ and $A^0 \rightarrow Z^0 Z^0$ can be written as the sum of
two parts. These are one loop decays that can have standard model fields
and supersymmetric fields in the loops. The standard model and
supersymmetric contributions do not mix due to their tensor structure. 
The supersymmetric diagrams can be broken into two form factors based on
gauge invariance, for the $A^0(p_5) \rightarrow \gamma(-Q^\mu)
Z^0(-p_3^\nu)$ decay we can write the one-loop $\gamma Z^0 A^0$ vertex
with squark loops as
\begin{equation}
i\Gamma_{\gamma, \textsc{susy}}^{\mu \nu} =
\eta^{\mu\nu}A^q_{\gamma} + Q^\nu p_3^\mu E^q_{\gamma}.
\end{equation}

The $Z^0 Z^0 A^0$ one-loop vertex will take the same form, but will have
different coefficients ($A^q_Z$ and $E^q_Z$). There is no term that is
proportional to the $\epsilon$ tensor in the squark loops because the
pseudoscalar does not couple to squarks with a $\gamma^5$. However, this
is the case when there are standard model fields in the loop. Therefore,
the same $\gamma Z^0 A^0$ vertex with quark loops can be written as
\begin{equation}
i\Gamma_{\gamma, \textsc{sm}}^{\mu \nu} =
\epsilon^{\mu \nu \alpha \beta} p_{3,\alpha} Q_\beta.
\end{equation}
When we try to interfere these two vertices it is now easy to see that
we find zero due to a repeated index in the $\epsilon$ tensor in the
first term and a term anti-symmetric in the indices multiplied by one
symmetric in the indices in the second term. Therefore, the decay widths
will be additive for these processes. This is also true for the
differential cross-section. Thus we can write
\begin{align}
\Gamma_{\textsc{tot}}  &=  \Gamma_{\textsc{sm}} + 
                           \Gamma_{\textsc{susy}}, \\
d\sigma_{\textsc{tot}} &= d\sigma_{\textsc{sm}} + 
                          d\sigma_{\textsc{susy}}.
\end{align}

This leads to some interesting phenomenology. It needs to be noted that
the pseudoscalar Higgs boson does not appear in the Standard Model, thus
the standard model contributions listed here are the contributions from
the standard model fields, in this case, the top and bottom quarks. Thus
deviations from the SM contributions do not tell us about the existence
of supersymmetry in nature, but they do tell us about the mixing in the
squark sector directly.

When we add all the squark loop diagrams we can determine the unknown
coefficients in our vertices. Thus, we can write
\begin{align} \nonumber
A^{q}_\gamma = \frac{4 ie^2\tilde{A}_q s_q c_q}{s_w c_w}
            &  \biggl\{
               s_q^2 T^3_q \bigl[ 
                 C_{24}(112) + C_{24}(221) - 2C_{24}(111) \\
            & - 2 Q_q [ B_0(21) - B_0(22) ]
                           \bigr]
               - Q_q^2 s_w^2 [ B_0(22) - B_0(11) ]
               \biggr\} - 
               \biggl\{ s_q \leftrightarrow c_q ;
                        1   \leftrightarrow 2
               \biggr\}, \\
E^q_\gamma = A^q_\gamma(B_0 \rightarrow 0 &; 
                        C_{24} \rightarrow C_{12} + C_{23} ),
\end{align}
for the $\gamma Z^0 A^0$ vertex and 
\begin{align} \nonumber
A^{q}_Z = \frac{-4 ie^2\tilde{A}_q s_q c_q}{s_w^2 c_w^2}
       &  \biggl\{
          c_q^4 (T^3_q)^2 \bigl[
                          C_{24}(121) + C_{24}(211) - 2C_{24}(111)
                          \bigr] \\ \nonumber
       &- c_q^2T^3_qQ_qs_w^2 \bigl[
                             C_{24}(122)+ C_{24}(121)+C_{24}(211)
                            +C_{24}(221)-4C_{24}(111)
                             \bigr] \\ \nonumber
       &+ 4 c_q^2 T^3_q \bigl[ B_0(12) - B_0(22) \bigr]
        - 2 Q_q s_w^2   \bigl[ B_0(22) - B_0(11) \bigr] \\ \nonumber
       &+ Q_q^2s_w^4C_{24}(111)
        - s_q^2c_q^2(T^3_q)^2 \bigl[
                              C_{24}(122)+C_{24}(121)
                             +C_{24}(112)+C_{24}(221)
                              \bigr]
          \biggr\} \\
       &- \biggl\{ s_q \leftrightarrow c_q ;
                   1   \leftrightarrow 2
          \biggr\}, \\
E^q_Z = A^q_Z(B_0 \rightarrow 0 &; C_{24} \rightarrow C_{12} + C_{23} )
\end{align}
for the $Z^0 Z^0 A^0$ vertex. The $C_{ij}$ functions are the usual
functions that appear in the Passarino-Veltman reduction
prescription\cite{Passarino:1978jh}. It is easy to see that these
expressions are finite. The $B_0$ functions appear in pairs with
opposite signs, canceling the $1/\epsilon$ poles which are independent
of their arguments. The $C_{24}$ functions also have argument
independent $1/\epsilon$ poles that all cancel in the expressions. The
same is true of the $C_{12}$ and $C_{13}$ functions. These functions can
be written out fully as
\begin{align}
C_{ij}(123) & \equiv C_{ij}(M_Z^2,Q^2,M_{A^0}^2 ; 
                            m_{\tilde{q}_1}^2 , 
                            m_{\tilde{q}_2}^2 , 
                            m_{\tilde{q}_3}^2 ) \\
B_0(12) & \equiv B_0(M_Z^2 ; m_{\tilde{q}_1}^2 , m_{\tilde{q}_2}^2 )
\end{align}
where $Q^2 = \{ 0,M_Z^2, s \}$ for the $A^0 \rightarrow \gamma Z^0$,
$A^0 \rightarrow Z^0Z^0$, and $e^+e^- \rightarrow Z^0 A^0$ processes
respectively. This completes the missing squark contributions to the 
decay widths and the differential cross-section for the \eeZA\ process.


\begin{thebibliography}{2005}

\bibitem{Gunion:1989we}
J.~F.~Gunion, H.~E.~Haber, G.~L.~Kane and S.~Dawson,
``The Higgs Hunter's Guide'', (Addison-Wesley, Reading, MA,
1990), Erratum ibid. [arXiv:hep-ph/9302272].

\bibitem{Carena:2002es}
M.~Carena and H.~E.~Haber,
Prog.\ Part.\ Nucl.\ Phys.\  {\bf 50}, 63 (2003)
[arXiv:hep-ph/0208209].

\bibitem{Heinemeyer:2004gx}
S.~Heinemeyer, W.~Hollik and G.~Weiglein,
arXiv:hep-ph/0412214.

\bibitem{Haber:1984rc}
H.~E.~Haber and G.~L.~Kane,
Phys.\ Rept.\  {\bf 117}, 75 (1985).

\bibitem{Harlander:2002vv}
R.~V.~Harlander and W.~B.~Kilgore,
JHEP {\bf 0210}, 017 (2002)
[arXiv:hep-ph/0208096].

\bibitem{Field:2002gt}
B.~Field,
Phys.\ Rev.\ D {\bf 66} (2002) 114007
[arXiv:hep-ph/0208262].

\bibitem{Field:2002pb}
B.~Field, J.~Smith, M.~E.~Tejeda-Yeomans and W.~L.~van Neerven,
Phys.\ Lett.\ B {\bf 551}, 137 (2003)
[arXiv:hep-ph/0210369].

\bibitem{Ravindran:2002dc}
V.~Ravindran, J.~Smith and W.~L.~Van Neerven,
Nucl.\ Phys.\ B {\bf 634}, 247 (2002)
[arXiv:hep-ph/0201114].

\bibitem{Field:2003yy}
B.~Field, S.~Dawson and J.~Smith,
Phys.\ Rev.\ D {\bf 69}, 074013 (2004)
[arXiv:hep-ph/0311199].

\bibitem{Field:2004tt}
B.~Field,
Phys.\ Rev.\ D {\bf 70}, 054008 (2004)
[arXiv:hep-ph/0405219].

\bibitem{Field:2004nc}
B.~Field,
arXiv:hep-ph/0407254.

\bibitem{Kao:1991xg}
C.~Kao,
Phys.\ Rev.\ D {\bf 46}, 4907 (1992).

\bibitem{Yin:2002sq}
J.~Yin, W.~G.~Ma, R.~Y.~Zhang and H.~S.~Hou,
Phys.\ Rev.\ D {\bf 66}, 095008 (2002).

\bibitem{Kao:2003jw}
C.~Kao, G.~Lovelace and L.~H.~Orr,
Phys.\ Lett.\ B {\bf 567}, 259 (2003)
[arXiv:hep-ph/0305028].

\bibitem{Kao:2004vp}
C.~Kao and S.~Sachithanandam,
arXiv:hep-ph/0411331.

\bibitem{Li:2005qn}
Q.~Li, C.~S.~Li, J.~J.~Liu, L.~G.~Jin and C.~P.~Yuan,
arXiv:hep-ph/0501070.

\bibitem{Barger:1993wt}
V.~D.~Barger, K.~m.~Cheung, A.~Djouadi, B.~A.~Kniehl and P.~M.~Zerwas,
Phys.\ Rev.\ D {\bf 49}, 79 (1994)
[arXiv:hep-ph/9306270].

\bibitem{Akeroyd:1999gu}
A.~G.~Akeroyd, A.~Arhrib and M.~Capdequi Peyranere,
Mod.\ Phys.\ Lett.\ A {\bf 14}, 2093 (1999)
[Erratum-ibid.\ A {\bf 17}, 373 (2002)]
[arXiv:hep-ph/9907542].

\bibitem{Akeroyd:2001ak}
A.~G.~Akeroyd, A.~Arhrib and M.~Capdequi Peyranere,
Phys.\ Rev.\ D {\bf 64}, 075007 (2001)
[Erratum-ibid.\ D {\bf 65}, 099903 (2002)]
[arXiv:hep-ph/0104243].

\bibitem{Farris:2002ny}
T.~Farris, J.~F.~Gunion and H.~E.~Logan,
in {\it Proc. of the APS/DPF/DPB Summer Study on the Future of Particle 
Physics (Snowmass 2001) } ed. N.~Graf,
eConf {\bf C010630}, P121 (2001)
[arXiv:hep-ph/0202087].

\bibitem{Arhrib:2002ti}
A.~Arhrib,
Phys.\ Rev.\ D {\bf 67}, 015003 (2003)
[arXiv:hep-ph/0207330].

\bibitem{Bartl:1997yd}
A.~Bartl, H.~Eberl, K.~Hidaka, T.~Kon, W.~Majerotto and Y.~Yamada,
Phys.\ Lett.\ B {\bf 402}, 303 (1997)
[arXiv:hep-ph/9701398].

\bibitem{Eberl:1999he}
H.~Eberl, K.~Hidaka, S.~Kraml, W.~Majerotto and Y.~Yamada,
Phys.\ Rev.\ D {\bf 62}, 055006 (2000)
[arXiv:hep-ph/9912463].

\bibitem{Accomando:1997wt}
E.~Accomando {\it et al.}  [ECFA/DESY LC Physics Working Group],
Phys.\ Rept.\  {\bf 299}, 1 (1998)
[arXiv:hep-ph/9705442].

\bibitem{Dawson:2004xz}
S.~Dawson and M.~Oreglia,
arXiv:hep-ph/0403015.

\bibitem{Eidelman:2004wy}
S.~Eidelman {\it et al.}  [Particle Data Group],
Phys.\ Lett.\ B {\bf 592}, 1 (2004).

\bibitem{Carena:1999py}
M.~Carena, D.~Garcia, U.~Nierste and C.~E.~M.~Wagner,
Nucl.\ Phys.\ B {\bf 577}, 88 (2000)
[arXiv:hep-ph/9912516].

\bibitem{Gunion:1991cw}
J.~F.~Gunion, H.~E.~Haber and C.~Kao,
Phys.\ Rev.\ D {\bf 46}, 2907 (1992).

\bibitem{Djouadi:1997yw}
A.~Djouadi, J.~Kalinowski and M.~Spira,
Comput.\ Phys.\ Commun.\  {\bf 108}, 56 (1998)
[arXiv:hep-ph/9704448].

\bibitem{Passarino:1978jh}
G.~Passarino and M.~J.~G.~Veltman,
Nucl.\ Phys.\ B {\bf 160}, 151 (1979).

\end{thebibliography}
\end{document}